\documentclass[11pt,a4paper]{article}
\usepackage{moriond}
\usepackage{amssymb,graphicx}

\def\Journal#1#2#3#4{{#1} {\bf #2}, #3 (#4)}

\def\NIMA{{\em Nucl.~Instrum.~Methods} A}

\def\PLB{{\em Phys.~Lett.}~B}
\def\PRL{\em Phys.~Rev.~Lett.}
\def\PRD{{\em Phys.~Rev.}~D}

\def\Astropart{{\em Astropart.~Phys.}}

\def\be{\begin{equation}}
\def\ee{\end{equation}}
\def\bea{\begin{eqnarray}}
\def\eea{\end{eqnarray}}

\begin{document}
\vspace*{4cm}
\title{DARK MATTER SEARCH WITH LIQUID NOBLE GASES}

\author{MARC SCHUMANN~\footnote{{\tt marc.schumann@physik.uzh.ch}}}
\address{Physik Institut, University of Zurich, Winterthurerstrasse 190, 8057 Zurich, Switzerland}

\maketitle

\abstracts{Dark matter detectors using the liquid noble gases xenon and argon as WIMP targets have evolved rapidly in the last decade and will continue to play a major role in the field. Due to the possibility to scale these detectors to larger masses relatively easily, noble liquids will likely be the first technology realizing a detector with a ton-scale target mass. In this article, we summarize the basic concepts of liquid noble gas dark matter detectors and review the current experimental status.}

\section{Introduction}

There is plenty of indirect cosmological evidence~\cite{ref::wimpcosmology} that the vast majority of the Universe's energy content is dark, 
with about 25\% being in the form of dark matter which builds large scale structures.
Another 70\% is the mysterious dark energy, responsible for the accelerated expansion of the Universe and only
about 5\% is ``ordinary'', baryonic matter which forms stars, planets, and eventually us.

There is no known particle in the Standard Model of Particle Physics which could be the dark matter particle,
neutrinos for example are too light and fast, hence the dark matter particle must be from new physics and is yet unknown. One of the most favorite candidates is the weakly interacting massive particle (WIMP)~\cite{ref::wimps}, which arises naturally in several extensions of the Standard Model, such as Supersymmetry, Universal Extra Dimensions, and little Higgs models. Many experiments aim at the direct detection of these particles by measuring nuclear recoils of target nuclei after they interact with a WIMP~\cite{ref::directdetect}. Sensitive detectors are placed in deep underground laboratories in order to fight backgrounds induced by cosmic rays and their daughter particles. The expected WIMP interaction rate is less than 1~event per kg of target material and year, and the featureless recoil spectrum is exponentially falling with typical energies of tens of~keV only. 

These experiments use different targets and detection methods, which all have different pros and cons. Liquid noble gases such as xenon and argon, but possibly also neon, have started to play an important role in the field since about a decade and are currently placing the most stringent limits on spin-independent WIMP-nucleon cross-sections over a large range of WIMP masses~\cite{ref::xe100run08,ref::xe10s2only}. This article gives a brief review on these detectors.

\section{Noble Gases as WIMP Targets}

The noble gases neon (Ne), argon (Ar), and xenon (Xe), which in liquid phase are all used or being considered as target materials for WIMP searches, have boiling points between 27.1~K (Ne) and 165.0~K (Xe), see Table~\ref{tab::gases}. This makes operation easier than for cryogenic detectors which have to be run at mK~temperatures. Xe and Ar can be even liquefied using liquid nitrogen. All three elements are excellent scintillators with very high light yields, and liquid xenon (LXe) and liquid argon (LAr) are very good ionizers as well, allowing for a direct measurement of the ionization signal induced by particle interactions. For this reason, mainly LAr and LXe are employed in current and future experiments, while neon is currently only considered as one option for one experiment (see CLEAN in Sect.~\ref{ref::experiments}). Hence, we restrict this summary to these two elements.

\begin{table}[tb]
\caption{\label{tab::gases} Selected properties of noble gases being used as WIMP targets. $W_{ph}$ and $W$ are the average energies to create a scintillation photon or an electron-ion pair. Numbers taken from~$^6$.}
\begin{center}
\begin{tabular}{|l|lll|}
\hline 
Element & Xenon & Argon & Neon \\ \hline
Atomic Number $Z$& 54 & 18 & 10 \\
Atomic mass $A$ & 131.3 & 40.0 & 20.2 \\
Boiling Point $T_b$ [K] &  165.0 & 87.3 & 27.1 \\
Liquid Density @ $T_b$ [g/cm$^3$] &  2.94 & 1.40 & 1.21 \\
Fraction in Earth's Atmosphere [ppm] & 0.09 & 9340 & 18.2 \\ \hline
Price & \$\$\$\$ & \$ & \$\$ \\
Scintillator & \checkmark & \checkmark & \checkmark \\
\ \ $W_{ph}$ ($\alpha,\beta$) [eV] & 17.9 / 21.6 & 27.1 / 24.4 & \\
\ \ Scintillation Wavelength [nm] & 178 & 128 & 78 \\ 
Ionizer & \checkmark & \checkmark & -- \\
\ \ $W$ ($E$ to generate e-ion pair) [eV] &  15.6 &  23.6 & \\
Experiments \footnotesize{[stopped, running, in preparation]} &  $\sim 5$ & $\sim 5$ & 1/2 \\
\hline
\end{tabular}
\end{center}
\end{table}

\begin{figure}[tb]
\begin{center}
\includegraphics*[width=0.7\columnwidth]{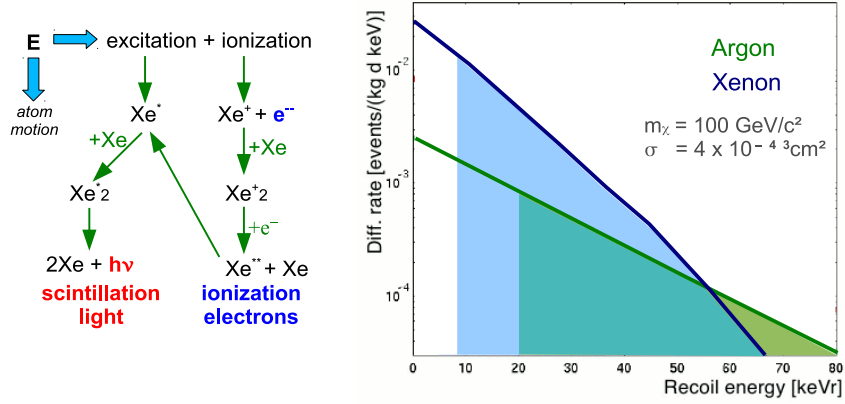}
\end{center}
\caption{\label{fig::scint} (left) Particle interactions excite and ionize the target (Xe in this example, but Ar works exactly the same way). Excited atoms Xe$^*$ combine with a neutral atom and form an excimer state Xe$^*_2$ which decays under the emission of scintillation light. If ionization electrons are not removed from the interaction site (e.g., by an electric field in a TPC), they eventually recombine and also produce scintillation light. Therefore, the light and the charge signal are anti-correlated. (right) Expected nuclear recoil spectra from interactions of a 100~GeV/c$^2$ WIMP with LXe and LAr, assuming a cross-section of $\sigma=10^{-43}$~cm$^2$. The expected rate is higher in LXe at low energies, but form factor suppressed at higher energies, which is not the case for LAr. A low detection threshold is therefore necessary if LXe is used. Experimentally achieved thresholds are indicated by the colored areas.}
\end{figure}

More material properties are summarized in Table~\ref{tab::gases}. In particular the possibility to build large, monolithic detectors make cryogenic noble liquids interesting for WIMP searches, as it is considered to be somewhat easier to scale these detectors up to the ton scale and beyond. Compared to the expensive Xe, the price of Ar is rather modest. However, while Xe is intrinsically clean from the radioactive point of view (there are no long-live Xe isotopes and contaminations of radioactive the $^{85}$Kr can be removed by cryogenic distillation), radioactive $^{39}$Ar is present in natural argon at the 1~Bq/kg level. This leads to background and pile-up problems. Finally, the wavelength of the scintillation light is at 178~nm for LXe which is observable with commercially available photocathodes, while LAr based detectors need to employ wavelength shifters (such as TPB) to detect the light.
Fig.~\ref{fig::scint} (left) shows how scintillation light is generated in liquid noble gases.  

The expected nuclear recoil energy $E_{nr}$ spectra from spin-independent WIMP-nucleon scattering interactions are featureless exponentials, and the interaction rate is expected to scale with $A^2$, see Fig.~\ref{fig::scint} (right). Therefore, the much heavier Xe is preferred. However, since the nucleus is also larger coherence is lost for large momentum transfers, leading to a form factor suppression of the rate at higher $E_{nr}$. A low detector threshold is therefore more mandatory for LXe. This is not the case for LAr, however, its overall interaction rate is always smaller. 

\paragraph{Background Discrimination}

If they exist, WIMPs are feebly interacting particles and their expected interaction rates are much lower than the omnipresent backgrounds from natural radioactivity and induced by cosmic rays. The latter can be decreased considerably (typically by factors $\sim10^{-6}$) by placing the experiments in deep underground laboratories, protected by several~km of rock overburden. Environmental radioactivity requires additional massive shields in which the detectors, built from selected radiopure materials, are being placed. The most critical background is from neutrons as these interact with the target nuclei and generate nuclear recoils, which makes them indistinguishable from WIMPs if they interact only once. However, neutrons have a rather large probability to interact several times inside a detector what allows for their rejection if these individual interactions can be resolved. The liquids, in particular LXe, have a rather high stopping power which can be used for self-shielding if only the inner part of the detector is selected for analysis (``fiducialization''). This, however, requires that the interaction vertex can be fairly well reconstructed in 3~dimensions.

The most abundant background for almost all dark matter experiments is from gamma and beta backgrounds which generate electronic recoils. These have a different energy loss d$E$/d$x$ compared to nuclear recoils leading to detectable differences in their signals which can be used for signal/background discrimination. The first possibility for noble liquids is the pulse shape of the scintillation signal. The excimers (see Fig.~\ref{fig::scint}) eventually emitting the light can be formed in singlet and triplet states which have different decay times. The individual population of the states depends on the particle interaction. In LAr, the lifetimes are about 3~order of magnitude different, with 0.005~$\mu$s and 1.6~$\mu$s for the singlet and triplet state, respectively, leading to a large slow component of the pulse for events from electronic recoil interactions. It has been demonstrated that this feature can be used to reject electronic recoils at the $3\times10^{-8}$ level~\cite{ref::LArDiscr}. However, such high levels are mandatory in order to cope with the huge background from $^{39}$Ar. With 4~ns and 22~ns, the singlet and triplet lifetimes are very similar in LXe and only very moderate rejection levels ($\sim0.1$) can be achieved~\cite{ref::LXeDiscr}, hence it is not used as default in any experiment.

If the charge and the light signal generated in an interaction are measured simultaneously for every event, one can exploit that the different d$E$/d$x$ for electronic recoil backgrounds and nuclear recoils signals produce a different charge/light ratio. The discrimination depends on the deposited energy and on the electric field strength applied to extract the charge signal. In LXe, it ranges between values of $5\times10^{-3}$ and $1\times10^{-4}$ at 50\% nuclear recoil acceptance~\cite{ref::xes2s1}. It is also used in LAr, however, its performance is much weaker than the pulse shape discrimination channel and would by itself not be sufficient to reduce the $^{39}$Ar background to the required low levels.

\section{Detector Concepts}

\begin{figure}[tb]
\begin{center}
\includegraphics*[width=0.8\columnwidth]{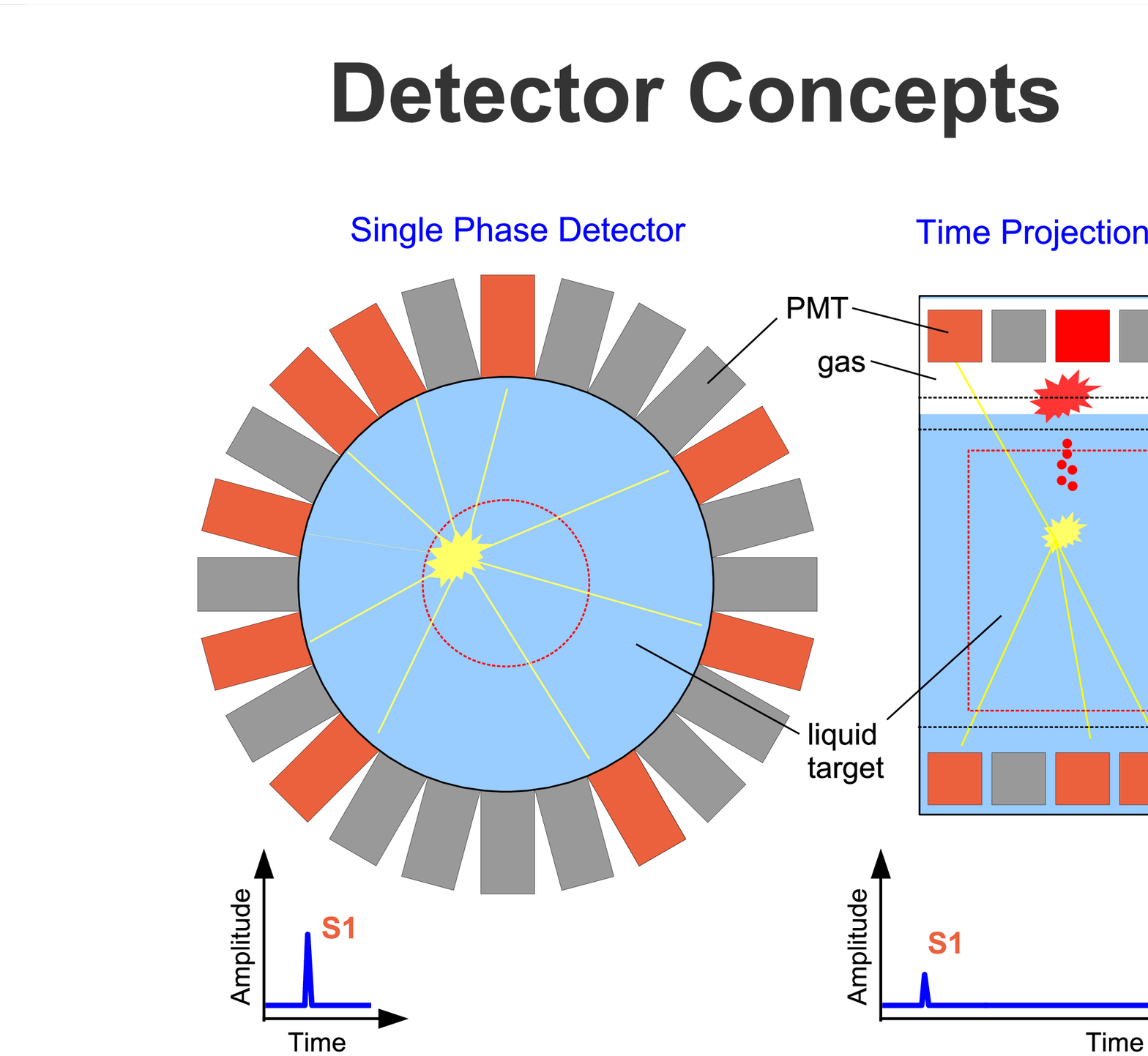}
\end{center}
\caption{\label{fig::concepts} The two detector concepts currently used for dark matter detectors based on liquid noble gases. (Left) Single phase detectors are essentially a large volume of a noble liquid which is viewed by many photosensors, usually PMTs, in order to detect the scintillation light S1. (Right) In a double phase detector the S1 signal is also detected by photosensors, but the ionization charge signal is measured as well since the detector is operated as a time projection chamber (TPC). An electric field across the target volume removes the ionization electrons from the interaction site and drifts them towards the gas phase on top of the liquid. The electrons are extracted into the gas and generate proportional scintillation light S2, which is registered time-delayed by the drift time. }
\end{figure}

Detectors using liquid noble gases as WIMP targets are currently operated by using two different concepts, which are illustrated in Fig.~\ref{fig::concepts} and are explained below.

\paragraph{Single Phase Detectors} 

These detectors are conceptually very simple devices in which a large volume of a liquid noble gas is viewed by as many light sensors (usually PMTs) as possible in order to reduce the detection threshold, see Fig.~\ref{fig::concepts} (left). Since only rather short scintillation light signals have to be detected, it also allows for rather high event rates since pile-up is almost no issue. The chosen geometry is usually spherical in order to exploit self shielding as much as possible. The $4\pi$~arrangement of the PMTs can be used for some rough event vertex reconstruction, with a resolution of typically several~cm. The reconstruction performance, however, depends on the number of detected photons and deteriorates close to threshold. Since only the light is detected, background discrimination via the charge/light ratio is not possible. Hence experiments have to rely on pulse shape discrimination or, in case of LXe, on almost perfect background reduction by shielding. For this reasons, most experiments will only use the innermost part of the detector as WIMP target and the outer part (which can be almost up to 90\% of the mass) as background shield.

\paragraph{Double Phase Detectors, Time Projection Chambers} 

Time projection chambers (TPCs), see Fig.~\ref{fig::concepts} (right), provide much better 3-dimensional vertex reconstruction, with demonstrated $z$-resolutions below 1~mm and a $xy$-resolution of $\sim 3$~mm~\cite{ref::xe100run08}. This is achieved by measuring the scintillation light and the ionization charge signal simultaneously. A particle interaction leads to scintillation and liberates ionization electrons which are removed from the interaction site by a strong electric field $E$ (``drift field'', typically around 1~kV/cm). The electrons drift towards the top of the cylindrical detector, where they are extracted into the gas phase above the liquid and generate a secondary light signal which is proportional to the charge~\cite{ref::tpc}. The lightpattern on the top PMT array is used to derive the $xy$-position and the time difference between light (S1) and charge (S2) signal to determine $z$. The excellent vertex detection capabilities allow for powerful background rejection via fiducialization and multi-scatter identification, accompanied by charge/light discrimination (plus pulse shape discrimination for LAr detectors). On the other hand, the optical coverage with photosensors is usually considerably smaller compared to single phase detectors, which might lead to an increased threshold. Additionally, one has to deal with the technical challenges related to the necessary high voltage system.

\section{Current Experiments using Noble Liquids}\label{ref::experiments}

In this section, we give a brief overview on most experimental efforts which currently employ
liquid noble gases as WIMP targets or which will use them in the near future. We have collected this
information to the best of our knowledge (using the experiment's presentations given recently~\cite{ref::ucladm}), it represents the status of May~2012. For space reasons, some projects have been omitted. 
 
\begin{figure}[h!]
\begin{center}
\includegraphics*[width=0.8\columnwidth]{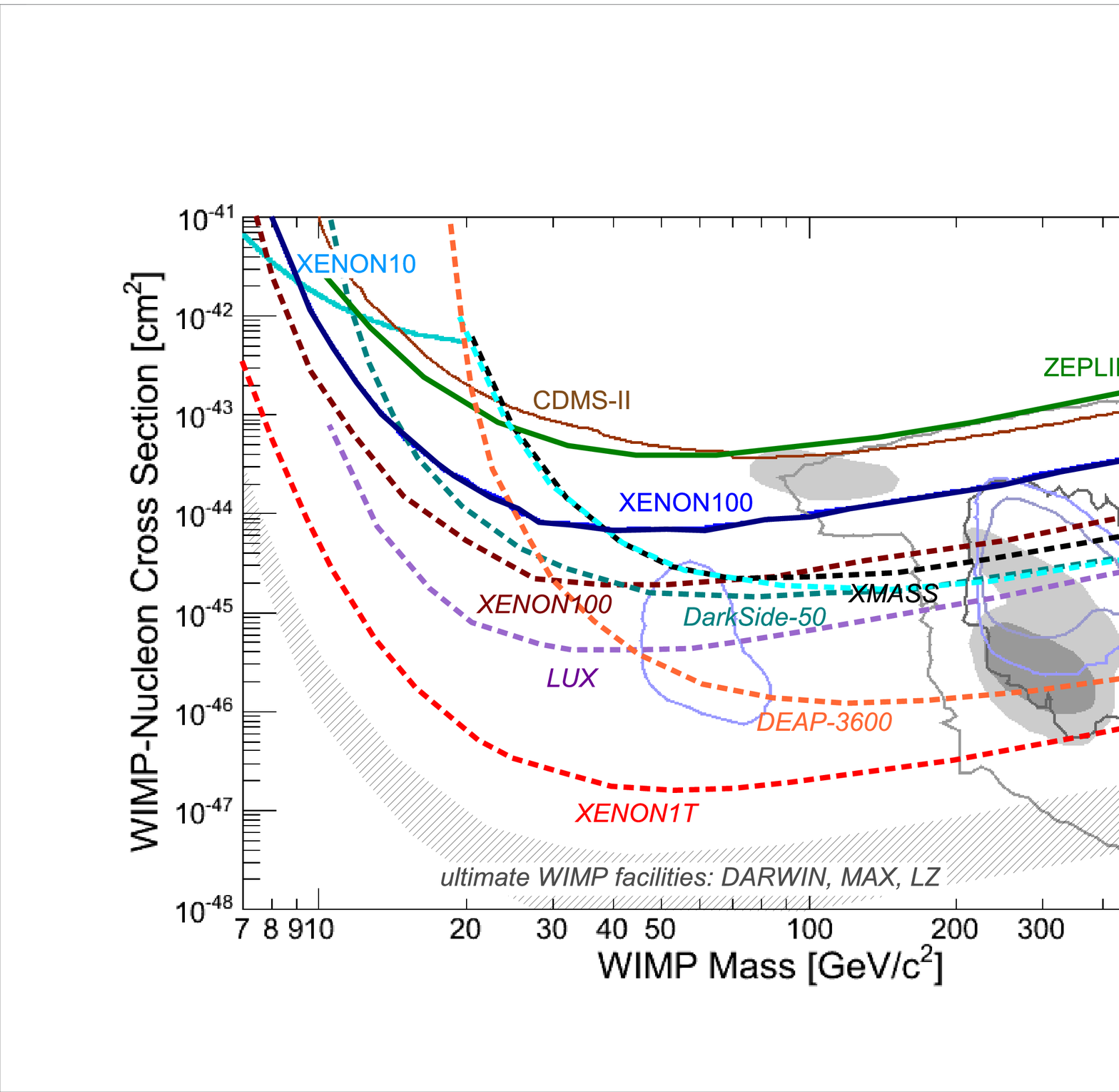}
\end{center}
\caption{\label{fig::limits} Achieved 90\% exclusion limits (solid lines) and projected sensitivities (dashed) of various dark matter projects using liquid noble gases (with the exception of CDMS-II). Shown is the spin-independent WIMP-nucleon scattering cross section vs.~the WIMP mass. Not all existing or planned experiments are shown, and most of the current experimental constraints are omitted. The closed areas indicate theoretically preferred SUSY regions~$^{12}$.}
\end{figure}

Experimentally achieved WIMP exclusion limits (at 90\% CL) are shown as solid lines in Fig.~\ref{fig::limits} for spin-independent WIMP-nucleon scattering interactions. Projected sensitivities are indicated by dashed lines. The most stringent limit to date comes from the XENON100 experiment~\cite{ref::xe100run08} excluding cross sections above $7.0\times10^{-45}$~cm$^2$ for $m_\chi=50$~GeV/$c^2$. Below $\sim 10$~GeV/$c^2$, the best limit is from XENON10~\cite{ref::xe10s2only}.

\paragraph{ZEPLIN-III} was a 12~kg double-phase LXe TPC out of which 5.1~kg were used as WIMP target. The experiment was installed in Boulby mine, UK. The extremely flat TPC geometry allowed for a very high drift field of 3-4~kV/cm and therefore a charge/light background rejection of $\sim1\times10^{-4}$. In the last science run from 2010-2011~\cite{ref::zeplin}, 8~events were observed in the predefined WIMP search region which was compatible with the background expectation and therefore led to an exclusion limit. With this result, the long history of ZEPLIN experiments has come to an end.

\paragraph{XENON100 and XENON1T} The current stage of the phased XENON program is XENON\-100, a double-phase LXe TPC with a total mass of 161~kg, located at Laboratori Nazionali del Gran Sasso (LNGS), Italy. 62~kg are inside the TPC and the remaining xenon surrounding the target in $4\pi$ is used as active veto. In the last science run~\cite{ref::xe100run08} of 100.9~days$\times$48~kg raw exposure, three events were observed, fully compatible with the expected background of $(1.8\pm0.6)$~events. A limit was placed which is currently setting the most stringent constraints for $m_\chi>10$~GeV/$c^2$. The results of a new dataset with about twice the exposure, a lower background, and a lower trigger threshold will be published soon.

The collaboration is already working on the next phase, XENON1T, which aims to explore cross sections down to $2\times10^{-47}$~cm$^2$ by 2017, after two years of data taking with a TPC of 1~ton LXe fiducial mass. XENON1T will be also installed at LNGS, inside a water shield of $\sim10$~m diameter which is operated as Cerenkov muon veto and will suppress ambient gamma radiation and neutrons.

\paragraph{XMASS} is a Japan-based single phase LXe detector, which aims for sensitivities around $2 \times 10^{-45}$~cm for $m_\chi = 100$~GeV/$c^2$~\cite{ref::xmass}. It employs a total of 800~kg LXe and uses about 100~kg as WIMP target. Since end of 2010 it is installed and running in Kamioka mine. A very high light yield has been achieved due to the large coverage with photosensors ($\sim60$\%). A first year of science data has been already collected, however, the collaboration has recently announced some issues with unexpected radioactive background from the PMTs~\cite{ref::xmassproblems}. XMASS is currently working to reduce the background.

\paragraph{LUX} is a double-phase LXe TPC which will be installed at the Sanford Underground Research Facility (SURF, USA). The detector employs a total of 350~kg of LXe and aims for a 100~kg fiducial mass for the WIMP search~\cite{ref::lux}. It is currently operated above ground to have a fully working detector once the underground space is ready for occupation. It has already demonstrated a rather high light yield and underground science is expected to start end of 2012 aiming at 300~days of data taking.

\paragraph{DarkSide} is a double-phase TPC which will use LAr as WIMP target~\cite{ref::darkside}. The goal for the next years is to build and operate DarkSide-50 with about 50~kg target mass. It will be located at LNGS (Italy), inside the the Borexino counting test facility (CTF), a large water tank which is currently being refurbished for this purpose. Inside the water shield, DarkSide will be surrounded by a spherical boron-loaded liquid scintillator neutron veto and it will use Ar which is depleted in $^{39}$Ar by a factor $\sim100$. Commissioning is scheduled for end of 2012, and two years of data taking is necessary to reach the final sensitivity around $10^{-45}$~cm$^2$.

\paragraph{ArDM} is a double-phase LAr detector~\cite{ref::ardm} which has been installed and commissioned at CERN and is currently being moved underground to the Canfranc laboratory (Spain). It employs a large target mass of 850~kg of LAr in a TPC of 120~cm height and 80~cm diameter. The collaboration is developing novel ways to deal with the technical challenges of multi-ton LAr/LXe detectors. The high voltage to bias the TPC, e.g., is generated next to the field cage in a Greinacher circuit and ArDM's final goal is to detect the charge signal with sub-mm precision in large micro-machined charge amplification detectors (large electron multipliers, LEMs).

\paragraph{DEAP-3600 and MiniCLEAN} is a large single phase detector using 3.6~tons of LAr, with about 1000~kg being used as WIMP target~\cite{ref::deap}. The LAr will be contained inside an acrylic vessel installed in a cryostat which itself is inside a water shield. Construction of the experiment is ongoing at SNOLab (Canada) and the first filling is expected around the end of 2013. The science goal is to reach the $10^{-46}$~cm$^2$ level after 3~years of operation. The large light collection in the single phase setup will allow for a very good rejection of electronic recoil background via pulse shape discrimination. 

The ``twin''-experiment MiniCLEAN~\cite{ref::miniclean} is being installed right next to DEAP-3600. With 150~kg LAr fiducial mass (500~kg total) it is considerably smaller, however, the experiment is being designed such that it can also be operated with liquid neon (LNe). Initially, this has been proposed in order to detect low energy neutrinos from the Sun and from supernovae~\cite{ref::clean}. However, if a signal is being seen in LAr it can be very useful to cross check this finding using the same detector (and the same systematics) with another target nucleus. MiniCLEAN is expected to run from end of 2012 to 2014.

\paragraph{Ultimate WIMP Facilities} Even though experimental results with ton-scale detectors have not been realized yet, several collaborations have already started to study the ultimate WIMP facilities which will explore the parameter space around $\sigma_\chi=10^{-48}$~cm$^2$, where neutrinos will be an homogeneously distributed irreducible background. The existing proposals DARWIN~\cite{ref::darwin}, MAX~\cite{ref::max}, and LZ~\cite{ref::lz} are all double phase TPCs. At the current stage, all these projects are just design and R\&D studies for multi-ton LXe and LAr detectors, which will likely not being built before 2020.

\section{Summary and Outlook} 

Many experiments aim to directly detect WIMP dark matter by searching for nuclear recoils from elastic WIMP collisions inside very sensitive detectors with ultra-low backgrounds. A large number of projects  employs the noble gases xenon or argon, cooled down and liquefied in order to obtain high-density targets. We have detailed why these elements are excellent WIMP targets and have explained the most common detector concepts. These are either single phase detectors measuring the scintillation light signal only, or double-phase detectors measuring the light and the charge signal (from ionization) in a TPC setup. 

At the time of writing, the most stringent exclusion limits for all WIMP masses are from LXe based detectors. We have presented the current status of more than 10~experiments using noble liquids which are all aiming to reach even higher sensitivities. Their goal is to explore new regions in the cross section vs.~mass parameter space (see Fig.~\ref{fig::limits}) and to finally detect the dark matter particle with detectors of 100-1000~kg target mass or even beyond.

\section*{Acknowledgments} We would like to thank the organizers of Rencontres de Moriond Cosmology 2012 
for their kind invitation to this great conference.



\section*{References}
\begin{thebibliography}{99}

\bibitem{ref::wimpcosmology}  N.~Jarosik {\it et al.}, {\it Astrophys.~J.~Suppl.}~{\bf 192}, 14 (2011); \\
K.~Nakamura {\it et al.} (Particle Data Group), {\it J.~Phys.~G} {\bf 37}, 075021 (2010).
\bibitem{ref::wimps} G.~Steigman and M.~S.~Turner, {\it Nucl.~Phys.~B}~{\bf 253}, 375 (1985); \\ G.~Jungman, M.~Kamionkowski, and K.~ Griest, {\it Phys.~Rept.}~{\bf 267}, 195 (1996).
\bibitem{ref::directdetect} M.~W.~Goodman and E.~Witten, \Journal{\PRD}{31}{3059}{1985}.
\bibitem{ref::xe100run08} E.~Aprile {\it et al.} (XENON100), \Journal{\PRL}{107}{131302}{2011}.
\bibitem{ref::xe10s2only} J.~Angle {\it et al.} (XENON10), \Journal{\PRL}{107}{051301}{2011}. 
\bibitem{ref::aprilebertone} E.~Aprile and L.~Baudis, in {\it Particle Dark Matter}, ed.~G.~Bertone (Cambridge University Press, 2010).
\bibitem{ref::LArDiscr} M.G.~Boulay {\it et al.}, arXiv:0904.2930; \\ H.~Lippincott {\it et al.} \Journal{PRC}{78}{035801}{2008}.
\bibitem{ref::LXeDiscr} J.~Kwong {\it et al.}, \Journal{\NIMA}{612}{328}{2010}; \\ K.~Ueshima {\it et al.} (XMASS), \Journal{NIMA}{659}{161}{2011}.
\bibitem{ref::xes2s1} E.~Aprile {\it et al.}, \Journal{\PRD}{72}{072006}{2005}; \\ V.~Lebedenko {\it et al}, \Journal{\PRD}{80}{052010}{2009}.
\bibitem{ref::tpc} B.~Dolgoshein, V.~Lebedenko, B.~Rodionov, {\it JETP Lett.}~{\bf 11}, 513 (1970); \\ A.~Bolozdynya {\it et al.}, {\it IEEE Trans.~Nucl.~Sci.}~{\bf 42}, 565 (1995).
\bibitem{ref::ucladm} Presentations given at UCLA Dark Matter 2012: \\ {\tt https://hepconf.physics.ucla.edu/dm12/agenda.html}.
\bibitem{ref::theory} A.~Fowlie {\it et al.}, \Journal{\PRD}{85}{075012}{2012}; \\ O.~Buchmueller {\it et al.}, arXiv:1112.3564; \\ 
C.~Strege {\it et al.}, {\it JCAP} {\bf 1203}, 030 (2012).
\bibitem{ref::zeplin} D.~Akimov {\it et al.} (ZEPLIN-III), \Journal{\PLB}{709}{14}{2012}.
\bibitem{ref::xmass} H.~Sekiya (XMASS), arXiv:1006.1473.
\bibitem{ref::xmassproblems} XMASS, presentation at the Japanese Physical Society (JPS) meeting, March 30, 2012.
\bibitem{ref::lux} D.S.~Akerib {\it et al.} (LUX), \Journal{\NIMA}{668}{1}{2012}.
\bibitem{ref::darkside} D.~Akimov {\it et al.} (DarkSide), arXiv:1204.6218.
\bibitem{ref::ardm} A.~Marchionni {\it et al.} (ArDM), arXiv:1012.5967.
\bibitem{ref::deap} M.G.~Boulay (DEAP-3600), arXiv:1203.0604.
\bibitem{ref::miniclean} A.~Hime (MiniCLEAN), arXiv:1110.1005.
\bibitem{ref::clean} D.N.~McKinsey, K.J.~Coakley, \Journal{\Astropart}{22}{355}{2005}. 
\bibitem{ref::darwin} L.~Baudis (DARWIN), arXiv:1201.2402; \\ M.~Schumann (DARWIN), arXiv:1111.6251.
\bibitem{ref::max} {\tt www.fnal.gov/pub/max}.
\bibitem{ref::lz} D.C.~Malling {\it et al.} (LZ), arXiv:1110.0103.
\end{thebibliography}
\end{document}